\documentclass[12pt,twoside,english]{article}
\usepackage[utf8]{inputenc}
\usepackage[nodisplayskipstretch]{setspace}
\usepackage{graphicx}
\graphicspath{ {./images/} }

\usepackage[paper=a4paper,dvips,top=1.5cm,left=1.5cm,right=1.5cm,
    foot=1cm,bottom=1.5cm]{geometry}

\usepackage{todonotes}          

\usepackage{mathptmx}
\usepackage[scaled=.90]{helvet}
\usepackage{courier}
\usepackage{bookmark}

\usepackage{fancyhdr}
\usepackage[utf8]{inputenc} 
\usepackage[english]{babel}
\usepackage{array}			 
\usepackage{graphicx}	                 
\usepackage{float}			 
\usepackage{color}                       
\usepackage{mdwlist}
\usepackage{tabularx}		         
\usepackage{dcolumn}	                 
\usepackage{url}	                 
\usepackage[perpage,para,symbol]{footmisc} 
\usepackage[binary-units=true]{siunitx} 

\usepackage[all]{hypcap}	 

\definecolor{darkblue}{rgb}{0.0,0.0,0.3} 
\definecolor{darkred}{rgb}{0.4,0.0,0.0}
\definecolor{red}{rgb}{0.7,0.0,0.0}
\definecolor{lightgrey}{rgb}{0.8,0.8,0.8} 
\definecolor{grey}{rgb}{0.6,0.6,0.6}
\definecolor{darkgrey}{rgb}{0.4,0.4,0.4}
\makeatletter
\renewcommand\paragraph{\@startsection{paragraph}{4}{\z@}%
  {-3.25ex\@plus -1ex \@minus -.2ex}%
  {1.5ex \@plus .2ex}%
  {\normalfont\normalsize\bfseries}}
\makeatother

\makeatletter
\renewcommand\subparagraph{\@startsection{subparagraph}{5}{\z@}%
  {-3.25ex\@plus -1ex \@minus -.2ex}%
  {1.5ex \@plus .2ex}%
  {\normalfont\normalsize\bfseries}}
\makeatother

\usepackage[acronym, nopostdot]{glossaries}
\glsdisablehyper
\makeglossaries

\newacronym{VM}{VM}{Virtual Machine}
\newacronym{MQTT}{MQTT}{Message Queuing Telemetry Transport}
\newacronym{AMQP}{AMQP}{Advanced Message Queuing Protocol}
\newacronym{GCP}{GCP}{Google Cloud Platform}
\newacronym{IoT}{IoT}{Internet of Things}
\newacronym{IIoT}{IIoT}{Industrial Internet of Things}
\newacronym{OS}{OS}{Operating System}
\newacronym{ms}{ms}{milliseconds}
\newacronym{KVM}{KVM}{(Kernel-based Virtual Machine)}
\newacronym{AWS}{AWS}{Amazon Web Services}
\newacronym{IaaS}{IaaS}{Infrastructure as a Service}
\newacronym{KB}{KB}{Kilobytes}

\title{Choosing an effective setup for stream processing }
\author{
        \textsc{Federico Ruilova,}
            \qquad
        \textsc{Aleksandar Yonchev}
        \mbox{}\\
        \normalsize
            \texttt{fedra@kth.se}
        \textbar{}
            \texttt{yonchev}
        \normalsize
            \texttt{@kth.se}
}
\date{\today}

\makeatletter
\let\ps@plain\ps@fancy 
\makeatother

\begin{document}

\maketitle

\begin{abstract}
\label{sec:abstract}


This project aims to study the feasibility and cost-effectiveness of using edge computing for stream data processing in the context of \gls{IoT} in manufacturing in Europe. Two scenarios were considered: using edge computing to reduce latency and using a popular public cloud provider. Both scenarios demonstrated high throughput, with the edge computing scenario slightly outperforming the public cloud scenario. The impact on resource utilization was also measured, with the edge node showing slightly lower resource usage than the cloud node. The experiment concluded that running the system at the edge is more cost-efficient, but only using any \gls{IaaS} provider acting as the infrastructure provider. \gls{IaaS} providers will be crucial in offering edge solutions and identifying geographical areas where regional data centers could be used as points of presence for low-latency applications.

\end{abstract}

\selectlanguage{english}
\tableofcontents

\newpage
\section{List of Acronyms and Abbreviations}
\label{list-of-acronyms-and-abbreviations}
\renewcommand{\glossarysection}[2][]{} 
\printglossary[type=\acronymtype,nonumberlist]

\clearpage
\section{Introduction}
\label{sect:introduction}

\subsection{Background}
\label{sec:background}

The number of devices connected to the Internet has been growing exponentially in the last decades \cite{iot_number_of_devices_statista}. Novelties in communication technologies such as the evolution of cellular networks into 5th generation networks with cloud computing, edge computing and a widely deployed network of high-speed fiber-optic infrastructure worldwide have opened many doors for new technologies to emerge. \newline

\noindent In its essence, edge computing or fog computing (the terms are used interchangeably \cite{shi_edge_2016}), serves the mission of bringing the cloud computing services closer to the client, by offloading some calculations to a nearby computing node. It is widely recommended for low latency and fast processing \cite{khan_edge_2019}, \cite{phan_dynamic_2021}, \cite{silva_investigating_2019}. However, there could be different approaches in terms of designing the most appropriate solution. There are three main architectures that are commonly used- a single edge-computing node, hybrid edge-cloud scenario and cloud-computing only.

\subsection{Literature review}
\label{sec:literature-review}
In \cite{silva_investigating_2019} an experiment is used where sample data is sent to the cloud, either directly or through an edge gateway. The two approaches are compared based on metrics like latency, throughput and available bandwidth. Different benefits of using edge computing are outlined at the end of the article. However, this experiment does not take into consideration the operating costs, neither the geographical location of the endpoints. In \cite{benchmarking_edge_cloud_hybrid} a similar experiment is run and there the latency, scalability and performance are compared between an edge-only, edge-cloud and cloud-only architectures. Similarly, strategies for selecting the best infrastructure by using an organization and management goal-based approach are recommended in \cite{vitali_special_2022}. Furthermore, there is a conducted research where the cloud-only scenario is used to evaluate the performance of transferring protocols for long distances \cite{almobayed_efficient_nodate}. 

The costs for using edge and cloud resources have been studied in other research projects. A detailed cost analysis on edge computing is performed in \cite{smart_edge_power}. Similarly, in \cite{cost_analysis_cloud_computing} a cost analysis is conducted over the usage of cloud computing resources. However, in \cite{ec2_vs_inhouse_cost_analysis} the Amazon EC2 service is compared to running in-house facilities for high-performance computing. We will refer to these resources when conducting the cost analysis.

The novelty of our approach comes from the cost analysis of running the different computing resources combined with the consideration of the geographical distances between the source of data and the computational resources in Europe. This project can serve as the basis for a broader research project which examines the demand for edge computing in relation to the requirements of the use case, the geographical distances (regardless of the continent) and the costs.

\subsection{Problem statement}
\label{sec:problem-statement}
\begin{figure}[!htb]
   \begin{minipage}{0.48\textwidth}
     \centering
     \includegraphics[width=.9\linewidth]{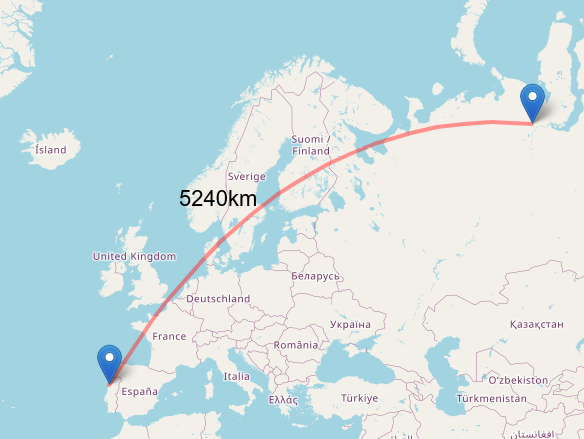}
     \caption{Distance between the westernmost and easternmost points in Europe}\label{Fig:distance-east-west}
   \end{minipage}\hfill
   \begin{minipage}{0.48\textwidth}
     \centering
     \includegraphics[width=.9\linewidth]{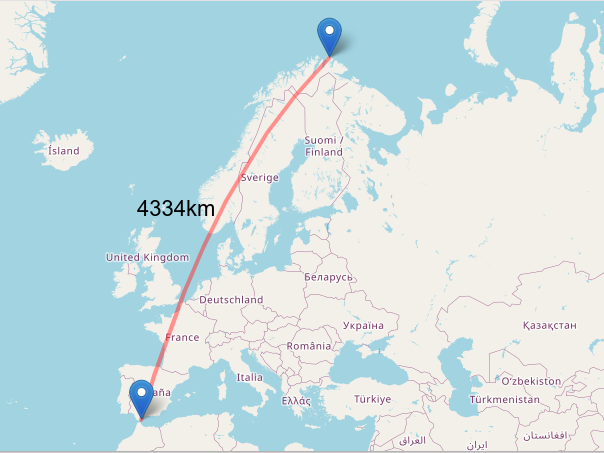}
     \caption{Distance between the northernmost and southernmost points in Europe}\label{Fig:distance-north-south}
   \end{minipage}
\end{figure}

The continent of Europe is the target of this research project because of its specific economical and geographical characteristics. Geographically speaking, Europe is the second-smallest continent on Earth, with a size of 10.18 km\textsuperscript{2} \cite{europe_wiki}. The distances between the extreme points in Europe are illustrated on Figures \ref{Fig:distance-east-west} and \ref{Fig:distance-north-south}.  second-smallest

\begin{figure}[H]
    \centering
    \includegraphics[width=250px]{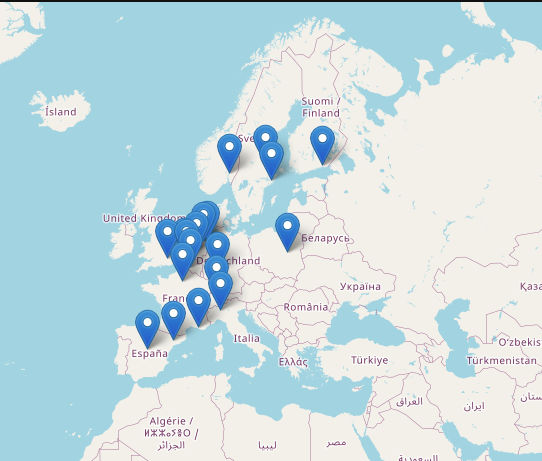}
    \caption{Public cloud provider data centers in Europe}
    \label{fig:data-center_Europe}
\end{figure}

On the economic side, Europe is a very industrialized and technologically developed area. Furthermore, all the major public cloud providers have one or more points of presence on the continent. Figure \ref{fig:data-center_Europe} shows the points of presence of Azure, \gls{AWS}, \gls{GCP}, IBM and Oracle in Europe. It can be clearly seen that there is a high density of public cloud data centers in the western, central and northern parts of Europe. However, there are no points of presence in Eastern Europe. This means that in Europe there still could be distances of more than 1000 kilometers between the client and the cloud provider.

According to \textit{Statista}, there are 2904 data centers in total in Europe \cite{data_centers_by_country_europe}. This means that there is a large amount of regional data centers which are not part of the big public cloud providers.

To summarize, three main observations are made: (a) cloud providers cover geographically main areas, (b) the major public cloud providers lack presence in Eastern Europe, (c) there is a high density of data centers within Europe, which do not belong to the popular public cloud providers. 

As mentioned earlier, edge computing brings cloud services closer to the client. This is done to ensure  ultra-low latencies of less than 10 \gls{ms} for the client. However, we will explore the applicability of edge computing when the distance between the client and the cloud is short. This leads to the main research question—"Is edge computing applicable in Europe?”.

Our hypothesis is that edge computing is not cost-efficient in Europe because of the short distances and easy access to public cloud services.





\newpage
\section{Method(s)}
\label{sec:method}

To answer the research question, a quantitative experiment was conducted. In this experiment, two \gls{IoT} environments were simulated – an edge and a cloud environment. 



To build these environments, the authors used three Virtual Machines (VMs)—one serving as the source of sensor data, another one serving as the edge node and a third node running in the cloud. 

In \glspl{IIoT}, it is common to have an edge processing node on-premises or in a nearby location. For the edge scenario, both machines were placed within the same data center, but they were not in the same network segment and had to use the public internet to communicate. The IaaS provider for this setup is Glesys, a Swedish provider with a point of presence in  Stockholm. 

In \ref{sec:problem-statement} it was mentioned that the major public cloud providers are still not present in Eastern Europe. For that reason, a more distant point of presence of \gls{GCP} was selected for the cloud scenario, to simulate a realistic scenario where the client is not in the center of Europe. In this scenario, the target node ran on the \gls{GCP}, specifically in their point of presence in St. Ghislain, Belgium. The distance between the source and the cloud nodes is 1,338 kilometers.

Through scripting, we could observe and gather information on latency, throughput, and resource utilization in each scenario.

\subsection{Technical specifications}
The “IoT Source” server runs on the KVM virtualization platform and has 2 CPU cores, 2048 MB of memory, and 20 GB of storage. It is also running the Ubuntu 22.04 LTS operating system. 

The “Edge-Node” server runs on the VMWare virtualization platform and has 2 CPU cores, 2048 MB of memory, and 30 GB of storage, and it is running the Ubuntu 22.04 LTS operating system. 

The “Cloud-Node” is a virtual machine with 2 virtual CPUs, 8 GB of memory and runs the Ubuntu 22.04 LTS operating system. It has a machine type of “n2-standard-2” and costs USD62.39 monthly. Google Compute Engine uses the KVM hypervisor to create and run VMs.

In our experiment we used the Eclipse Mosquitto message broker which implements the \gls{MQTT} protocol for stream processing. Mosquitto provides a light-weight communication between the devices using a publishing/subscribe messaging model and is a perfect match for \glspl{IIoT} scenarios \cite{eclipse_mosquitto} \cite{mqtt_implementation_power_plant_monitoring}. It also provides the ability to subscribe to multiple topics, authentication, encrypted communication and more. Furthermore, we used the Eclipse Paho client libraries for Mosquitto because the broker has only a command-line interface. Paho gave us the ability to manipulate the flow of \gls{MQTT} messages using Python code \cite{eclipse_paho}. For this experiment, we created 2 python scripts—one running in the data source and the second one running in the target, which is either tpublishing or the cloud node. The first script generates exemplary sensor data and sends it to the target, while the second captures the incoming messages, measures the latency for each message, and outputs the throughput and resource utilization at the end of the data transfer.



The remote brokers at the edge and the cloud subscribed to three topics called “sensor1”, “sensor2” and “sensor3”. We ran the scripts with different configurations to measure the different variables. Furthermore, we used a standard bandwidth configuration in both scenarios with a 100 Mbps connection.

First, we ran the experiment for 900 seconds (or 15 minutes) with a 1-second interval between each message transmission. The messages were randomly assigned to each sensor. This way, we measured the latency between the communicating sides.

Second, we simulated 1000 data transfer iterations at the IoT source and on each iteration 200 messages were published to the target.  The experiment ran for 1000 seconds, and each message was 10\gls{KB} in size. In this scenario, the interval between the messages was set to 0, but there was 1 second interval after each iteration.

For analyzing the data, we used Python because it is well suited for data analysis and scripting. The libraries we used are Matplotlib and NumPy to create the different plots (e.g. Figure \ref{fig:latency-distribution}) and find patterns in the data. Furthermore, for analyzing the distances and the position of the public cloud data centers, we used JavaScript with the open-source library Leaflet, which is perfect for creating interactive maps \cite{library_leaflet}. Leaflet uses the open-source geographic database OpenStreetMap \cite{wiki_openstreetmap}.

\newpage
\section{Results}
\label{sec:results}

\subsection{Latency}
\begin{figure}[H]
    \centering
    \includegraphics[width=300px]{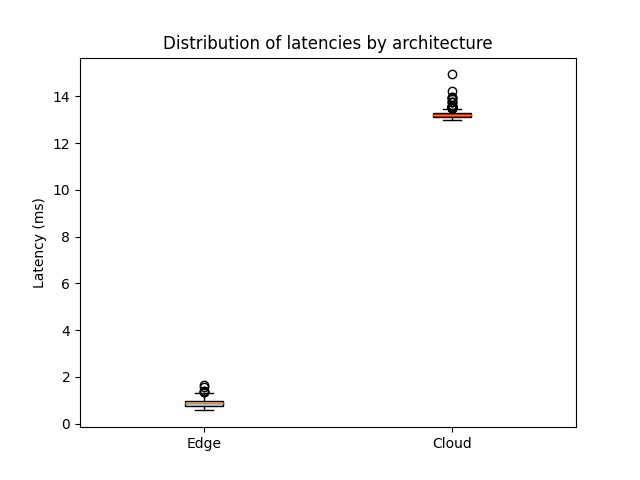}
    \caption{Distribution of latencies for the different architectures}
    \label{fig:latency-distribution}
\end{figure}

The normal distribution of the latencies measured through the experiment is visualized on Figure \ref{fig:latency-distribution}. In the first scenario, the median of the values is 0.87 \gls{ms}. The maximum latency recorded is 1.65 \gls{ms}. In the other scenario, however, the median is 13.19 \gls{ms}. The maximum recorded value is 14.94 \gls{ms}. 

\subsection{Throughput and Resource utilization}
For Scenario 1 (IoT Source to Edge Node), the authors measured the resource utilization impact on the IoT Source with an average CPU usage of 40.24 percent, an average memory usage of 18.87 MB, and an average running time of 1094 seconds\ref{performace source}. The performance impact on the Edge node was also measured, with an average CPU usage of 42.67 percent, an average memory usage of 21.41 MB, and an average running time of 1002 seconds\ref{Performance Impact Nodes}.

The system's Throughput in Scenario 1 (IoT Source to Edge Node) was approximately 3780 messages per second, equivalent to approximately 3.03 Mbps \ref{throughput table}. The authors calculated the throughput by dividing the total data transferred (378.0 MB) by the total time elapsed (1000 seconds) and multiplying the result by 8 to convert it to bits per second:

\begin{equation} \label{eq:1}
Throughput (Mbps) = (Total \; data \; transferred (MB) / Total \; time \; elapsed (seconds)) * 8 bits/byte = 
\end{equation}
\begin{equation} \label{eq:2}
    (378.0 MB / 1000 seconds) * 8 bits/byte = 3.03 Mbps
\end{equation}

For Scenario 2 (IoT Source to Cloud), the performance impact on the IoT Source had an average CPU usage of 41.34 percent, an average memory usage of 20.09 MB, and an average running time of 1001 seconds \ref{performace source}. The performance impact on the Cloud node was also measured, with an average CPU usage of 77.87 percent, an average memory usage of 16.95 MB, and an average running time of 1032 seconds \ref{Performance Impact Nodes}.

The system's Throughput in Scenario 2 (IoT Source to Cloud) was approximately 3820 messages per second, equivalent to approximately 3.06 Mbps \ref{throughput table}. The authors calculated the throughput with the same method mentioned in \ref{eq:1} and \ref{eq:2}.


\begin{table}[!ht]
    \centering
    \begin{tabular}{|l|l|}
    \hline
        \textbf{Scenario} & \textbf{Throughput (Mbps)} \\ \hline
        \textbf{IoT Source to Edge Node} & 3.03 \\ \hline
        \textbf{IoT Source to Cloud} & 3.06 \\ \hline
    \end{tabular}
    \caption{Throughput measured on each scenario}
    \label{throughput table}
\end{table}

\begin{table}[!ht]
    \centering
    \begin{tabular}{|l|l|l|}
    \hline
        \textbf{Scenario} & \textbf{CPU Usage} & \textbf{Memory Usage} \\ \hline
        IoT Source (Scenario 1) & 41\% & 20 MB \\ \hline
        IoT Source (Scenario 2) & 41.34\% & 20.09 MB \\ \hline
    \end{tabular}
     \caption{Comparison of system performance, IoT source in Scenarios 1 and 2}
    \label{performace source}
\end{table}

\begin{table}[!ht]
    \centering
    \begin{tabular}{|l|l|l|}
    \hline
        \textbf{Node}  & \textbf{CPU Usage} & \textbf{Memory Usage}\\ \hline
        \textbf{Edge Node (Scenario 1)}  & 42.67\% & 21.41 MB \\ \hline
        \textbf{Cloud Node (Scenario 2)}  & 77.87\% & 16.95 MB \\ \hline
    \end{tabular}
    \caption{Comparison of system performance, Edge vs. Cloud}
    \label{Performance Impact Nodes}
\end{table}



\subsection{Cost}
 As presented in \ref{Table of Costs} “IoT Source” costs 215,60 SEK/mo, or approximately €19.18 (using the same exchange rate). “Edge-Node” costs 402,88 SEK/mo, or approximately €35.94 (using an exchange rate of 1 SEK = €0.0886354 EUR).
According to the provider, the price difference lies in the virtualization platform VMWare compared to KVM.

The “Cloud-Node” cost was €44.15 (using an exchange rate of 1 EUR = €1.06449 USD) The "Cloud-Node” cost was €44.15 (using an exchange rate of 1 EUR = €1.06449 USD).

\begin{table}[!ht]
    \centering
    \begin{tabular}{|l|l|l|l|l|}
    \hline
        \textbf{Computational Node} & \textbf{Cost (USD/mo)} & \textbf{Cost (SEK/mo)} & \textbf{Cost (EUR/mo)} \\ \hline
        \textbf{IoT Source} & - & 215.60 & 19.18  \\ \hline
        \textbf{Edge-Node} & - & 402.88 & 35.94  \\ \hline
        \textbf{Cloud-Node} & 44.15 & - & 47.42  \\ \hline
    \end{tabular}
    \caption{Comparison of the costs for the different nodes}
    \label{Table of Costs}
\end{table}

\newpage
\section{Discussion}
\label{sec:Discussion}

Looking at the measured latencies, it is realistic to experience such low values (median of 0.87 \gls{ms}) for the source-edge environment, as both \glspl{VM} are positioned in the same data center in Stockholm. In the other scenario, the median latency measured is 13.19 \gls{ms}. This is a reasonable value, considering the 1,338 km distance between the hosts.

While there are different scenarios in IoT streaming platforms or other stream processing use cases, we have identified an important player in the cloud computing ecosystem, the IaaS providers, which do not own the infrastructure but offer “public-cloud” alike products such as VMs and make use of regional data centers to operate and offer their services. 

We have also noticed by conducting the experiment that it is more cost-efficient to run the infrastructure with an IaaS provider that uses KVM virtualization. There could be different providers offering their clouds on the top of similar virtualization platforms as the ones used by the major Public cloud providers.

\subsection{Conclusion}

Based on the observations, both scenarios demonstrated high throughput, with a slightly higher throughput in Scenario 2 (IoT Source to Cloud). 

These high throughput scenarios are likely due to the efficient design and implementation of the system,  as well as favorable network conditions and workload.


In both scenarios, the performance impact on the IoT Source was relatively similar, with an average CPU usage of around 41 percent and an average memory usage of around 20 MB. 

The performance impact on the Edge node and Cloud node was also measured, with the Edge node having slightly lower resource utilization compared to the Cloud node. It is important to note that the throughput of a system can be affected by various factors, including the hardware and software used, network conditions, and workload. 

We could replicate a low-latency setup not running on the cloud with a nearby presence without incurring acquisition costs. We achieved it by using the services of a \gls{IaaS} provider with the desired point of presence. 

Our hypothesis was proven wrong. Under the experiment's circumstances, we conclude that running the system at the edge is more cost-efficient but without incurring the cost of acquisition or ownership, meaning that an IaaS provider needs to act as the infrastructure provider. The point of processing needs to be identified by the system architects without relying upon only in major public cloud providers.

IaaS Providers will be essential in offering edge solutions and identifying geographical areas where regional data centers could be used as points of presence for low-latency applications. 

\subsection{Future work}
The research project could be extended by comparing a hybrid architecture, consisting of an edge and a cloud computing node. In addition, the experiment could be opened widely by simulating a real-world scenario where Apache Flink is used for processing the data stream. Real IoT devices, like Raspberry Pi or Arduino nodes, could also be used in the future.


Further experimentation and analysis are necessary to fully understand the factors influencing the system's performance. In this experiment, we compared KVM and VMWare, but were unable to see if that affected the slight difference in throughput in favor of the KVM platform at the cloud node.

Another consideration for the future would be to increase the available bandwidth. Currently, there are \glspl{VM} in \gls{GCP} specifically designed for streaming applications. We could utilize such a \gls{VM} and increase the bandwidth on the source and edge nodes, to see if this will provide new findings. However, because of the economical aspect, we couldn't afford to use the best-performing resources in this project. 

\newpage
\bibliography{Choosing-an-effective-setup-for-stream-processing}

\begin{thebibliography}{10}
\providecommand{\url}[1]{#1}
\csname url@samestyle\endcsname
\providecommand{\newblock}{\relax}
\providecommand{\bibinfo}[2]{#2}
\providecommand{\BIBentrySTDinterwordspacing}{\spaceskip=0pt\relax}
\providecommand{\BIBentryALTinterwordstretchfactor}{4}
\providecommand{\BIBentryALTinterwordspacing}{\spaceskip=\fontdimen2\font plus
\BIBentryALTinterwordstretchfactor\fontdimen3\font minus
  \fontdimen4\font\relax}
\providecommand{\BIBforeignlanguage}[2]{{%
\expandafter\ifx\csname l@#1\endcsname\relax
\typeout{** WARNING: IEEEtran.bst: No hyphenation pattern has been}%
\typeout{** loaded for the language `#1'. Using the pattern for}%
\typeout{** the default language instead.}%
\else
\language=\csname l@#1\endcsname
\fi
#2}}
\providecommand{\BIBdecl}{\relax}
\BIBdecl

\bibitem{iot_number_of_devices_statista}
\BIBentryALTinterwordspacing
``Edge computing deployments by region 2028.'' [Online]. Available:
  \url{https://www.statista.com/statistics/1104059/worldwide-edge-computing-infrastructure-region/}
\BIBentrySTDinterwordspacing

\bibitem{shi_edge_2016}
\BIBentryALTinterwordspacing
W.~Shi, J.~Cao, Q.~Zhang, Y.~Li, and L.~Xu, ``\BIBforeignlanguage{en}{Edge
  {Computing}: {Vision} and {Challenges}},'' \emph{\BIBforeignlanguage{en}{IEEE
  Internet of Things Journal}}, vol.~3, no.~5, pp. 637--646, Oct. 2016. doi:
  10.1109/JIOT.2016.2579198. [Online]. Available:
  \url{http://ieeexplore.ieee.org/document/7488250/}
\BIBentrySTDinterwordspacing

\bibitem{khan_edge_2019}
\BIBentryALTinterwordspacing
W.~Z. Khan, E.~Ahmed, S.~Hakak, I.~Yaqoob, and A.~Ahmed,
  ``\BIBforeignlanguage{en}{Edge computing: {A} survey},''
  \emph{\BIBforeignlanguage{en}{Future Generation Computer Systems}}, vol.~97,
  pp. 219--235, Aug. 2019. doi: 10.1016/j.future.2019.02.050. [Online].
  Available:
  \url{https://linkinghub.elsevier.com/retrieve/pii/S0167739X18319903}
\BIBentrySTDinterwordspacing

\bibitem{phan_dynamic_2021}
\BIBentryALTinterwordspacing
L.-A. Phan, D.-T. Nguyen, M.~Lee, D.-H. Park, and T.~Kim,
  ``\BIBforeignlanguage{en}{Dynamic fog-to-fog offloading in {SDN}-based fog
  computing systems},'' \emph{\BIBforeignlanguage{en}{Future Generation
  Computer Systems}}, vol. 117, pp. 486--497, Apr. 2021. doi:
  10.1016/j.future.2020.12.021. [Online]. Available:
  \url{https://www.sciencedirect.com/science/article/pii/S0167739X20330831}
\BIBentrySTDinterwordspacing

\bibitem{silva_investigating_2019}
\BIBentryALTinterwordspacing
P.~Silva, A.~Costan, and G.~Antoniu, ``\BIBforeignlanguage{en}{Investigating
  {Edge} vs. {Cloud} {Computing} {Trade}-offs for {Stream} {Processing}},'' in
  \emph{\BIBforeignlanguage{en}{2019 {IEEE} {International} {Conference} on
  {Big} {Data} ({Big} {Data})}}.\hskip 1em plus 0.5em minus 0.4em\relax Los
  Angeles, CA, USA: IEEE, Dec. 2019. doi: 10.1109/BigData47090.2019.9006139.
  ISBN 978-1-72810-858-2 pp. 469--474. [Online]. Available:
  \url{https://ieeexplore.ieee.org/document/9006139/}
\BIBentrySTDinterwordspacing

\bibitem{benchmarking_edge_cloud_hybrid}
\BIBentryALTinterwordspacing
F.~Carpio, M.~Delgado, and A.~Jukan, ``\BIBforeignlanguage{en}{Engineering and
  {Experimentally} {Benchmarking} a {Container}-based {Edge} {Computing}
  {System}},'' in \emph{\BIBforeignlanguage{en}{{ICC} 2020 - 2020 {IEEE}
  {International} {Conference} on {Communications} ({ICC})}}.\hskip 1em plus
  0.5em minus 0.4em\relax Dublin, Ireland: IEEE, Jun. 2020. doi:
  10.1109/ICC40277.2020.9148636. ISBN 978-1-72815-089-5 pp. 1--6. [Online].
  Available: \url{https://ieeexplore.ieee.org/document/9148636/}
\BIBentrySTDinterwordspacing

\bibitem{vitali_special_2022}
\BIBentryALTinterwordspacing
M.~Vitali, P.~Plebani, D.~Bermbach, and E.~Elmroth,
  ``\BIBforeignlanguage{en}{Special issue on co-design of data and computation
  management in {Fog} {Computing}},'' \emph{\BIBforeignlanguage{en}{Future
  Generation Computer Systems}}, vol. 129, pp. 423--424, Apr. 2022. doi:
  10.1016/j.future.2021.11.001. [Online]. Available:
  \url{https://www.sciencedirect.com/science/article/pii/S0167739X21004301}
\BIBentrySTDinterwordspacing

\bibitem{almobayed_efficient_nodate}
F.~AlMobayed, ``\BIBforeignlanguage{en}{Efficient {High} {Performance}
  {Protocols} {For} {Long} {Distance} {Big} {Data} {File} {Transfer}},'' p.
  134.

\bibitem{smart_edge_power}
A.~Mehta and L.~Eleftheriadis, ``Smart {Edge} {Power} {Management} to {Improve}
  {Availability} and {Cost}-efficiency of {Edge} {Cloud},'' in \emph{2022
  {IEEE} 15th {International} {Conference} on {Cloud} {Computing} ({CLOUD})},
  Jul. 2022. doi: 10.1109/CLOUD55607.2022.00032 pp. 125--133, iSSN: 2159-6190.

\bibitem{cost_analysis_cloud_computing}
X.~Li, Y.~Li, T.~Liu, J.~Qiu, and F.~Wang, ``The {Method} and {Tool} of {Cost}
  {Analysis} for {Cloud} {Computing},'' in \emph{2009 {IEEE} {International}
  {Conference} on {Cloud} {Computing}}, Sep. 2009. doi: 10.1109/CLOUD.2009.84
  pp. 93--100, iSSN: 2159-6190.

\bibitem{ec2_vs_inhouse_cost_analysis}
J.~Emeras, S.~Varrette, V.~Plugaru, and P.~Bouvry, ``Amazon elastic compute
  cloud (ec2) versus in-house hpc platform: A cost analysis,'' \emph{IEEE
  Transactions on Cloud Computing}, vol.~7, no.~2, pp. 456--468, 2019. doi:
  10.1109/TCC.2016.2628371

\bibitem{europe_wiki}
\BIBentryALTinterwordspacing
``\BIBforeignlanguage{en}{Europe},'' Jan. 2023, page Version ID: 1131266536.
  [Online]. Available:
  \url{https://en.wikipedia.org/w/index.php?title=Europe&oldid=1131266536}
\BIBentrySTDinterwordspacing

\bibitem{data_centers_by_country_europe}
\BIBentryALTinterwordspacing
``\BIBforeignlanguage{en}{Data centers by country in {Europe} 2022}.''
  [Online]. Available:
  \url{https://www.statista.com/statistics/878621/european-data-centers-by-country/}
\BIBentrySTDinterwordspacing

\bibitem{eclipse_mosquitto}
\BIBentryALTinterwordspacing
``\BIBforeignlanguage{en}{Eclipse {Mosquitto}},'' Jan. 2018. [Online].
  Available: \url{https://mosquitto.org/}
\BIBentrySTDinterwordspacing

\bibitem{mqtt_implementation_power_plant_monitoring}
\BIBentryALTinterwordspacing
{Hema}, ``\BIBforeignlanguage{en-US}{{MQTT} {Implementation} on {Power} {Plant}
  {Monitoring} - {IoT} {Use} case},'' Sep. 2020. [Online]. Available:
  \url{https://www.bevywise.com/blog/iot-success-stories-mqtt-broker-celikler-holding/}
\BIBentrySTDinterwordspacing

\bibitem{eclipse_paho}
\BIBentryALTinterwordspacing
I.~Craggs, ``\BIBforeignlanguage{en}{Eclipse {Paho} {\textbar} {The} {Eclipse}
  {Foundation}}.'' [Online]. Available: \url{https://www.eclipse.org/paho/}
\BIBentrySTDinterwordspacing

\bibitem{library_leaflet}
\BIBentryALTinterwordspacing
``\BIBforeignlanguage{en}{Leaflet — an open-source {JavaScript} library for
  interactive maps}.'' [Online]. Available: \url{https://leafletjs.com/}
\BIBentrySTDinterwordspacing

\bibitem{wiki_openstreetmap}
\BIBentryALTinterwordspacing
``\BIBforeignlanguage{en}{{OpenStreetMap}},'' Jan. 2023, page Version ID:
  1132335746. [Online]. Available:
  \url{https://en.wikipedia.org/w/index.php?title=OpenStreetMap&oldid=1132335746}
\BIBentrySTDinterwordspacing

\end{thebibliography}
\bibliographystyle{myIEEEtran}
\appendix

\end{document}